\documentclass[%
 reprint,
 amsmath,amssymb,
 aps,
]{revtex4-1}

\usepackage{graphicx}
\usepackage{dcolumn}
\usepackage{bm}
\usepackage{dsfont}

\usepackage{color}
\usepackage{ulem}

\begin{document}

\preprint{APS/123-QED}

\title{Comment on some particular phases of supersymmetric QCD}

\author{Renata Jora
	$^{\it \bf a}$~\footnote[1]{Email:
		rjora@theory.nipne.ro}}

\affiliation{$^{\bf \it a}$ National Institute of Physics and Nuclear Engineering PO Box MG-6, Bucharest-Magurele, Romania}

\begin{abstract}

 We study two phases of supersymmetric QCD  associated to the singularities of the renormalization constant of the fermion mass operator. We find a behavior of the potential between two charges for the phases of interest that it is consistent with Seiberg results on these aspects based on a different method.
\end{abstract}
\maketitle

The study of the phase diagram of  a gauge theory is of extreme relevance for its nonperturbative behavior. In \cite{Seiberg1}, \cite{Seiberg2} Seiberg elucidated the main aspects of the phase diagram of supersymmetric QCD using the powerful tools of holomorphicity and duality. Some properties of the associated phases for both QCD and  supersymmetric QCD were analyzed from the perspective of renormalization group equations   in \cite{Jora}.

In this work we will discuss two particular phases of supersymmetric QCD based on the singularities of the renormalization constant of the fermion mass operator $Z_2$. It turns out that those cases for which we can gather information are the free electric and the free magnetic phases already introduced by Seiberg for the same number of flavors.

The Lagrangian for supersymmetric QCD in terms of the holomorphic coupling has the expression:
\begin{eqnarray}
&&{\cal L}_h=\frac{1}{16}\int d^2\theta W^a(V_h)W^a(V_h)+h.c. +
\nonumber\\
&&\int D^4 \theta \sum_I\Phi_i^{\dagger}\exp[2V_h^i]\Phi_i,
\label{holo999}
\end{eqnarray}
where i represents the chiral multiplet $i$ and $V_h^i=V_h^aT^a_i$ where $T^a$ are the generators of the group in the fundamental representation of the chiral multiplet.

A transformation to the canonical couplings by the change of variable $V_h=g_cV_c$ \cite{Murayama} leads to the canonical Lagrangian with the right kinetic terms for the supermultiplets.  The corresponding beta function in the canonical coupling is the NSVZ beta function \cite{NSVZ}:
\begin{eqnarray}
\beta(g_c)=-\frac{g_c^3}{16\pi^2}\frac{3N-N_f+N_f\gamma(g_c)}{1-N\frac{g^2_c}{8\pi^2}},
\label{betaufnc665}
\end{eqnarray}
where $\gamma(g_c)$ is the anomalous dimension of the chiral multiplet mass operator which at one loop has the expression:
\begin{eqnarray}
\gamma(g_c)=\frac{ d \ln Z_2}{d \ln(\mu)}=-\frac{g_c^2}{8\pi^2}\frac{N^2-1}{N}.
\label{anom885774}
\end{eqnarray}

The connection between the holomorphic coupling and the canonical one is given by \cite{Murayama}:
\begin{eqnarray}
\frac{1}{g_h^2}=\frac{1}{g_c^2}+\frac{N}{8\pi^2}\ln(g_c^2)-\frac{N_f}{8\pi^2}\ln Z_2,
\label{connection5664}
\end{eqnarray}
where $Z_2$ is the anomalous dimension of the fermion mass operator.

It is known that the first two orders coefficients of a beta function and the first order coefficient of the anomalous dimensions are renormalization scheme independent. In consequence we may consider only the first order coefficient of the anomalous dimension relevant. In this case one can write:
\begin{eqnarray}
\frac{\partial  \ln Z_2}{\partial g_c}=\frac{\partial \ln Z_2}{ \partial \ln(\mu)}\frac{\partial \ln(\mu)}{\partial g_c}=\frac{\gamma (g_c)}{\beta(g_c)}.
\label{res77566499}
\end{eqnarray}
and integrate $d\ln Z_2$ using the expressions in Eqs. (\ref{betaufnc665}) and (\ref{anom885774}) to obtain:
\begin{eqnarray}
&&\ln Z_2={\rm const}+\frac{N}{3N-N_f}\ln(g_c^2)+
\nonumber\\
&&\frac{N(3N-2N_f)}{N_f(3N-N_f)}\ln[N_f-3N+NN_f\frac{g_c^2}{8\pi^2}].
\label{integr4665779}
\end{eqnarray}

In the following we shall discuss the singularities associated with the expression in Eq. (\ref{integr4665779}).

It is evident that $\ln Z_2$ diverges for $N_f=3N$. One can estimate the factor in front for $N_f=3N$:
\begin{eqnarray}
&&\ln Z_2={\rm const}+ \frac{1}{3N-N_f}\ln\Bigg[ g^{2N}(NN_f\frac{g^2}{8\pi^2})^{-N}\Bigg]=
\nonumber\\
&&-\frac{N}{3N-N_f}\ln[\frac{NN_f}{8\pi^2}],
\label{partic6577466}
\end{eqnarray}
to determine that in the Large N limit $\frac{NN_f}{8\pi^2}>1$ one has:

1) $N_f\approx 3N$ but $N_f$ slightly larger that $3N$  leads to $\ln Z_2=\infty$.

2) $N_f\approx 3N$ but $N_f$ slightly lower than $3N$ leads to $\ln Z_2=-\infty$.

Knowing that the holomorphic beta function runs at one loop one can write:
\begin{eqnarray}
\frac{1}{g_h^2}=\frac{3N-N_f}{8\pi^2}\ln[\frac{\mu}{\Lambda}],
\label{holomorphic77566}
\end{eqnarray}
where $\Lambda$ is the supersymmetric QCD scale and $\mu$ the running scale. For a fixed $\frac{1}{g_h^2}$  for $N_f\rightarrow 3N (N_f<3N)$ then $\mu\approx\infty$. For $N_f\rightarrow 3N (N_f>3N)$ then $\mu\approx0$. Consequently for $\mu\approx 0$, $\ln Z_2=\infty$ and for $\mu\approx\infty$, $\ln Z_2=-\infty$. Eq. (\ref{connection5664}) yields for  $3N=N_f$:
\begin{eqnarray}
0=\frac{1}{g_c^2}+\frac{N}{8\pi^2}\ln(g_c^2)-\frac{N_f}{8\pi^2}\ln Z_2.
\label{res66464}
\end{eqnarray}
Noting that $\frac{1}{g_c^2}=\frac{1}{Z_Ag_{c0}^2}$ where $g_{c0}$ is the bare coupling constant we observe that according to our previous arguments:
$Z_A\rightarrow 0_{+}$ for small momenta and $Z_A\rightarrow 0_{-}$ for large moments.  Since the actual on-shell-gluon propagator (one can consider gluino with the same result) has the expression:
\begin{eqnarray}
{\rm Gluon\,\,propagator}\approx\frac{1}{p^2}Z_A,
\label{res64663}
\end{eqnarray}
it is clear that in order for that to have the correct behavior in both low energy and large energy regime it  must be of the form:
\begin{eqnarray}
{\rm Gluon\,\,propagator}\approx\frac{1}{p^2\ln[\frac{\Lambda^2}{p^2}]}.
\label{propertygluon64665774}
\end{eqnarray}
But the potential between two charges is the Fourier transform in the three dimensional space of the expression in Eq. (\ref{propertygluon64665774}):
\begin{eqnarray}
&&V(r)\approx \int d^3\vec{p} \frac{1}{p^2\ln[\frac{\Lambda^2}{p^2}]}\exp[irp]\approx
\nonumber\\
&&\frac{1}{r}\frac{1}{\ln(\Lambda r)}\int d^3\vec{p}' \frac{1}{p^{\prime2}\ln[\frac{\Lambda^2}{p^{\prime 2}}]}\exp[ip] \approx\frac{1}{r}\frac{1}{\ln(\Lambda r)},
\label{finalres666477}
\end{eqnarray}
where in the last line we made the change of variables $pr=p'$.

Note that according to Seiberg classification \cite{Seiberg2} the potential in Eq. (\ref{finalres666477}) corresponds to the free electric phase. On the other hand using different arguments Seiberg found that the case $N_f=3N$ is associated to the free electric phase. Our findings thus confirm Seiberg's results.

The second singularity of interest in Eq. (\ref{integr4665779}) is when the logarithm diverges i.e. when:
\begin{eqnarray}
&&N_f-3N+NN_f\frac{g_c^2}{8\pi^2}=0
\nonumber\\
&&\frac{Ng_c^2}{8\pi^2}=\frac{3N-N_f}{N_f}
\label{secresuly7777}
\end{eqnarray}

Then indifferent of the coupling constant for $N_f\leq\frac{3N}{2}$ $\ln Z_2=-\infty$ and for $N_f\geq\frac{3N}{2}$ $\ln Z_2=+\infty$. The case $N_f=\frac{3N}{2}$ is  the only one we can discuss.  According to Eq. (\ref{connection5664}) then one can write:
\begin{eqnarray}
\frac{1}{g_h^2}={\rm const\,\,finite}-\frac{N_f}{8\pi^2}\ln (Z_2).
\label{eval775664}
\end{eqnarray}

The correct renormalization of the holomorphic coupling is $\frac{1}{g_h^2}=\frac{Z_A}{g_{h0}^2}$. Based on our previous arguments we determine that  for $N_f\leq\frac{3N}{2}$  $Z_A=\infty$, for $N_f\geq\frac{3N}{2}$
$Z_A=-\infty$. But,
\begin{eqnarray}
\frac{1}{g_h^2}=\frac{3N-N_f}{8\pi^2}\ln[\frac{\mu}{\Lambda}],
\label{res553442}
\end{eqnarray}
which means that for $\mu$ small $\frac{1}{g_h^2}\approx -\infty$ and for $\mu$ large $\frac{1}{g_h^2}=\infty$. Of course $\frac{1}{g_h^2}$ should be regarded in the asymptotic limit and it is assumed that may take any value. Also it is important to observe that $Z_A$ has the same behavior as $\frac{1}{g_h^2}$

The only possible gluon propagator compatible with this behavior is :
\begin{eqnarray}
{\rm Gluon\,\,propagator}\approx\frac{1}{p^2}\ln[\frac{p^2}{\Lambda^2}],
\label{finalpropoag5546}
\end{eqnarray}
which further leads to a potential:
\begin{eqnarray}
V(r)\approx\frac{1}{r}\ln(\Lambda r).
\label{finlares664774}
\end{eqnarray}

According to Seiberg classification this case corresponds to the free magnetic phase and again  our result agrees with Seiberg's results which state that for $N_f=\frac{3N}{2}$ the supersymmetric QCD is in the free magnetic phase.

In summary we analyzed two particular phase of supersymmetric QCD related with the number of flavors at which singularities in the renormalization constant of the fermions mass operator occur. It turns out that our study leads directly to a potential between two charges corresponding to the free electric and free magnetic phases respectively.  Our results agree exactly with Seiberg findings for $N_f=3N$ and $N_f=\frac{3N}{2}$. The method might shed some lights on other nonperturbative aspects.

\end{document}